\pdfoutput=0

\documentclass[twocolumn,a4paper]{article}
 \usepackage{graphicx}
 \usepackage{amssymb}
 \usepackage{amsthm}
 \usepackage{lineno}

\title{First Application of Pulse-Shape Analysis to Silicon Micro-Strip Detectors\\
~\\
\small{M.~von~Schmid$^{a,}$\footnote{Corresponding author at: Institut f\"{u}r Kernphysik, Technische Universit\"{a}t
Darmstadt, Schlossgartenstr. 9, D-64289 Darmstadt, Germany. e-Mail: schmid@ikp.tu-darmstadt.de}~, P.~Egelhof$^b$, V.~Eremin$^c$, R.~Gernh\"{a}user$^d$, T.~Kr\"{o}ll$^a$, M.~Mutterer$^{b,a}$, N.~Pietralla$^a$, B.~Streicher$^b$, M.~Weber$^d$}\\
~\\
\footnotesize{$^a$Institut f\"{u}r Kernphysik, Technische Universit\"{a}t Darmstadt, Schlossgartenstr. 9, D-64289 Darmstadt, Germany}\\
\footnotesize{$^b$GSI Helmholtzzentrum f\"{u}r Schwerionenforschung GmbH, Planckstr. 1, D-64291 Darmstadt, Germany}\\
\footnotesize{$^c$Physical - Technical Institute of Russian Academy of Sciences, 26 Polytechnicheskaya str., 194021 St. Petersburg, Russia}\\
\footnotesize{$^d$Physik-Department E12, Technische Universit\"{a}t M\"{u}nchen, James-Franck-Str., D-85748 Garching, Germany}}
\author{}
\date{}

\begin{document}

\maketitle

\begin{abstract}
The method of pulse-shape analysis (PSA) for particle identification (PID) was applied to a double-sided silicon strip detector (DSSD) with a strip pitch of $300$~$\mu$m. We present the results of test measurements with particles from the reactions of a $70$~MeV $^{12}$C beam impinging on a mylar target. Good separation between protons and alpha particles down to $3$~MeV has been obtained when excluding the interstrip events of the DSSD from the analysis.

\end{abstract}

\section{Introduction}
\label{introduction}

The identification of charged particles by nuclear charge and mass is a key demand in nuclear spectroscopy. A traditional method for identification of nuclear charges is based on energy loss measurements in a stack of two or more detectors, for example made of silicon detectors. The $\Delta$E detector introduces an energy threshold for particle identification and add an energy straggling, since the particle needs to punch through the first layer of the stack. If this detector is made of silicon, there exists a practical trade-off between minimal thickness and sensitive area of the detector: Thin silicon detectors (of the order of tens of micro\-meter) may provide low energy thresholds and acceptable amount of energy straggling, but are very limited in size and, on the contrary, larger detectors have thicknesses in the order of $100$~$\mu$m. During the last years pulse-shape analysis has become a suitable alternative for particle identification, delivering equivalent or even better results than traditional energy-loss measurements. This method is based on the fact that the pulse shape of the current signal, induced on the contacts of a silicon detector, contains information about nuclear charge and partly even mass of the particle. Two effects mainly define the time evolution of the current signal (see [1 -- 9] and references therein): on the one hand the signal is formed by superposition of the signals induced by both, electrons and holes. As the mobility of electrons is about three times larger than that of the holes, the shape of the signal depends on the depth where the charge carriers have been created. The linear energy transfer of the particle in matter depends on the particle's velocity and strongly on its nuclear charge Z. Therefore, the pulse shape will also be sensitive on these parameters. On the other hand the particle creates a plasma column along its trajectory in the detector's sensitive volume which shields the external electrical field so that charge carriers can be separated on the surface of the column only, resulting in a delayed charge collection. The delay time is called ``plasma erosion time'' and depends on the ionisation density and, therefore, again on the deposited energy and the type of the particle.

Already in 1963 Ammerlaan et al.\ \cite{Ammerlaan:1963} introduced an experimental method for discriminating particles with lithium drifted silicon detectors of $2200~\mu$m to $4400~\mu$m thickness exploiting the pulse shapes. Discrimination between alpha particles and deuterons was demonstrated in an energy range from $8$~MeV to $26$~MeV. It was already proposed that better separation could be obtained by injecting the particles from the rear side of the detector (non-junction side, n-side) where the effects due to varying penetration depths and plasma delay act coherently. Later, England et al.\ \cite{England:1989} used rear-side injection with a surface-barrier detector to perform pulse-shape analysis. They exploited the thin insensitive window layers of surface-barrier detectors. Progress in this field was achieved by using the fast output of a wide-band hybrid preamplifier. Pausch et al.\ published several experimental results (\cite{Pausch:1992}, \cite{Pausch:1994}) from pulse-shape analysis, which finally led them to develope a theoretical model \cite{Pausch:1994sim} for describing the pulse shape of a silicon detector taking into account the plasma effects. With this model they were able to explain the results of former pulse-shape analysis experiments and to improve the method for further experiments \cite{Pausch:1995}. Recently, Mutterer et al.\ \cite{Mutterer:2000} presented impressive results from pulse shape analysis by using surface-barrier detectors made of very homogeneous neutron-transmutation-doped (n-TD) silicon. They demonstrated good Z separation down to low energies equivalent to a penetration depth of $20~\mu$m in silicon. The effects of crystal-orientation effects (channeling) on the pulse-shape analysis were studied carefully by Bardelli et al.\ in \cite{Bardelli:2009}. These studies showed a non negligible impact of the crystal orientation on the pulse-shape analysis. In summary, the following features can be named as being most important for good performance of pulse-shape analysis:
\begin{itemize}
	\item homogeneous resistivity profile of the detector
	\item orientation of the silicon crystal with respect to the front surface of
the detector
	\item thin insensitive layers on the n-side to allow reversed operation of the detector
	\item wide-band and low-noise electronics to process the pulse shape
\end{itemize}
In recent publications from Bardelli et al.\ \cite{Bardelli:2004} and Barlini et al.\ \cite{Barlini:2009} digital sampling techniques were successfully utilized for pulse-shape analysis experiments. The authors outlined the high potential of digital pulse-shape analysis for example to decrease the lower energy threshold of the particle identification.

Segmented silicon detectors allow position sensitive measurements. The substantial possibilities of modern planar technology enable the production of a wide varity of segmented detectors for special purposes. Most segmented detectors, especially highly segmented detectors, are used mainly for particle tracking, but also spectroscopy is possible. So far, pulse-shape analysis with strip detectors has not been investigated often. Lu et al.\ \cite{Lu:2001} used radially segmented silicon detectors with a rather wide pitch of $8$~mm for pulse-shape analysis in combination with time-of-flight measurement. Also in combination with time-of-flight measurements, Alderighi et al.\ \cite{Alderighi:2004} used silicon detectors divided into 2 segments with a total sensitive area of $20$~cm$^2$ for particle identification. Another work utilizing a single-sided segmented silicon detector with 4 pads of $10~\times~10$~mm$^2$ each was published by John et al.\ \cite{John:2009}. Finally recent work with single sided strip detectors was done by Bardayan et al.\ \cite{Bardayan:2009}.
	\begin{figure}[t]
  	\centering
  	\includegraphics{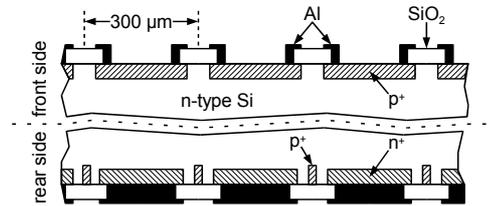}
  	\caption{Scheme of the used DSSD (strips at the rear side are displayed rotated by 90$^{\circ}$), see text for details.}
  	\label{fig:dssd}
	\end{figure}
	
In our work we will discuss a first approach of applying pulse-shape analysis to highly segmented double-sided silicon strip detectors (DSSD) \cite{mvs}. As their name implies they are segmented on both sides into strips (perpendicular to each other) allowing pixel-like two-dimensional position measurements. The work is motivated by the planed EXL experiment \cite{exl} which is a part of the NuSTAR project at the future FAIR facillity. The goal of EXL is the investigation of elastic and inelastic scattering as well as direct reactions (transfer and charge exchange) with exotic beams within the planed NESR storage ring. The reactions will be performed in inverse kinematics on light nuclei utilizing an internal gas jet or droplet target. The pulse-shape analysis with DSSDs offers some general advantages compared to other techniques and is particularly suited to meet the requirements of the EXL project. In the first instance, experimental situations could be addressed where neither $\Delta$E/E detector arrays nor time-of-flight measurements are suitable and a position sensitive measurement is needed. In addition, a position sensitive measurement also overcomes the homogeneity problems with using large area detectors for PSA. Especially larger detectors tend to show more inhomogeneity of the resistivity profile than smaller ones, resulting in a position dependence of the pulse-shape analysis performance. For the small area of one ``pixel'' the detector material can be expected to be sufficiently homogeneous and pulse-shape analysis can be done precisely pixel per pixel in order to be insensitive to global changes in detector homogeneity. On the other hand, it is an open question wether properties only present in segmented detectors, like distorted electrical fields between two adjacent strips or interstrip events, could affect proper particle identification.

	\begin{figure}[t]
  	\centering
  	\includegraphics{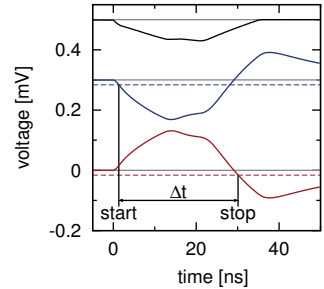}
  	\caption{Top curve: simulated pulse shape of a $5{.}5$~MeV alpha particle impinging the n-side of an silicon detector (bias voltage: $80$~V, depletion voltage: $50$~V); middle curve: shaped signal (amplified, differentiated: $20$~ns, integrated: $2$~ns) giving the start signal; bottom curve: same as above but inverted, giving the stop signal; dashed lines indicate LED trigger levels.}
  	\label{fig:simpulses}
	\end{figure}
	
In the following, we will present results of the pulse-shape analysis with a DSSD segmented into $16$ strips each side. Test measurements were performed at the Tandem-van-de-Graaff accelerator of the Maier-Leibnitz-Laboratorium, Garching. Mainly light particles up to $^4$He from the reactions of a $70$~MeV $^{12}$C beam impinging on a mylar target have been investigated. Results of these measurements will be presented.

\section{Experiment}
\label{experiment}
\subsection{Detector}
As already mentioned, we used a DSSD with $16$ strips on each side and a strip pitch of $300~\mu$m. It has outer dimensions of $7{.}1\times7{.}1$ mm$^2$ and is made of $300~\mu$m thick n-type silicon. The detector was developed and produced in cooperation with the Physical-Technical Institute, St.\ Petersburg, Russia as a small prototype for the EXL experiment. The strips of the junction side (p-side) are the boron implanted lines separated by an oxide layer. The strips of the rear side (n-side) are phosphorus implanted regions insulated by the p$^+$-implanted lines. Interstrip gaps are $15~\mu$m wide at the p-side and $65~\mu$m at the n-side. A schematic view of the detector is shown in figure \ref{fig:dssd}. This strip pattern is also representative for the larger EXL prototype detectors ($21\times21$~mm$^2$) investigated so far \cite{Streicher:2008}. The detector bias can be applied through the innermost of the 5 guard rings surrounding the detector (punch through biasing). The depletion voltage of this detector is about $50$~V and it was operated, slightly over-biased, between $60$~V and $80$~V.

\subsection{Electronics setup}
As the current signal is induced on both sides of the detector electrodes, the electronics setup for PSA was connected only to one side, whereas the other side was read out with standard electronics for the energy measurement. Energy signals from both, p- and n-side, are used for the discrimination of interstrip events. These are events, which cause signal sharing among two or more strips at either side of the detector. The method we used for PSA is similar to the one Mutterer et al.\ successfully used in \cite{Mutterer:2000}. The preamplifier in our setup was the same modified CSTA2. The CSTA2 (development of electronics laboratory of IKP at TU~Darmstadt \cite{Bonnes}) is a wide-band hybrid preamplifier with a slow output for energy measurements and a fast timing output. With the applied modification, the fast timing signal represents a good approximation of the current signal induced in the detector. This signal is amplified and shaped with a TFA2000-6 timing filter amplifier, which is also a development of IKP at TU~Darmstadt.

	\begin{figure}[t]
  	\centering
  	\includegraphics{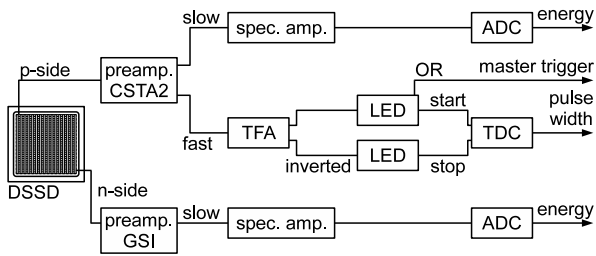}
  	\caption{Block diagram of the electronics used in the experiment, for details see text.}
  	\label{fig:electronicsetup}
	\end{figure}
	
\begin{figure*}[htb]
 \centering
 \includegraphics{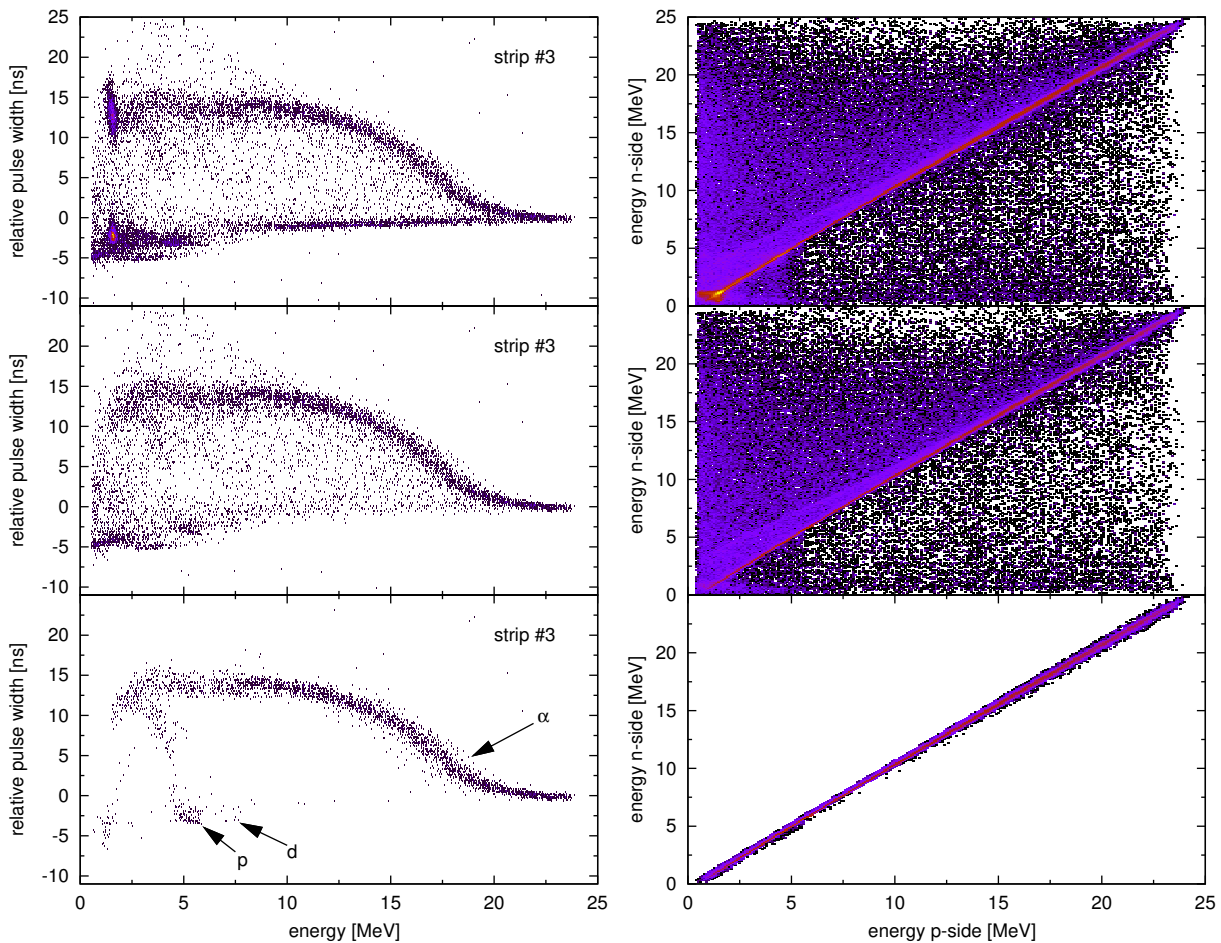}
 \caption{Left side: energy versus relative pulse width (offset normalized to $23$~MeV alpha-particle energy) exemplary for strip \#3; right side: energy correlation between p-side and n-side of the DSSD; conditions applied in the analysis (\textit{from top to bottom}): all events, only particles stopped in the DSSD, stopped particles without interstrip events.}
 \label{fig:TvsE}
\end{figure*}	
Figure \ref{fig:simpulses} (top curve) shows the simulation of a pulse shape induced by a $5{.}5$~MeV alpha particle (like from a $^{241}$Am alpha source) penetrating the n-side of a silicon detector (bias voltage: $80$~V, depletion voltage: $50$~V). The software code \cite{Evsenin} is based on a model by Pausch et al.\ \cite{Pausch:1994sim}. An alpha particle with such energy penetrates roughly $20~\mu$m into silicon and therefore produces a relatively slow pulse when stopped so close underneath the surface of the rear side (n$^+$-layer). To illustrate the principle of pulse-shape analysis, the simulated signal is amplified, differentiated with a time constant of $20$~ns and integrated with an integration time of $2$~ns (middle curve in figure~\ref{fig:simpulses}), the same way as it was done in the experiment. The time constants are chosen in such a way that an overshoot is produced for all pulses expected. Therefore, the time constant of the differentiation should be in the order of the smallest pulse width. The time constant of the integration should be set smaller than the rise time of the fastest pulses. The actual values have to be fine tuned with the oscilloscope as they depend in practice also on cable lentghs, stray capacities etc. which influence the shape of the signal. By using a Leading Edge Discriminator (LED) with a low threshold close to the noise level (indicated by a dashed line in figure \ref{fig:simpulses}), one can derive a start signal from the pulse. The same but inverted signal (bottom curve in figure \ref{fig:simpulses}) is used to get a stop signal by triggering with another LED on the undershoot, which is a result of the differentiation. The measured time difference between start and stop signal, digitized in a TDC, represents an approximation of the pulse width, which is our first observable for PSA. The second observable, the energy, is measured with the energy output of the CSTA2 preamplifier on the p-side and GSI developed hybrid preamplifiers DC coupled to the n-side. Amplifiers and ADCs are standard electronics. An electronic block scheme exemplary for one channel of each side is given in figure \ref{fig:electronicsetup}.

\subsection{Setup at the tandem accelarator at MLL}

We tested the particle identification with particles emitted following compound reactions of a $^{12}$C beam at $70$~MeV impinging on a $2~\mu$m thick mylar target. By placing an absorber foil ($100~\mu$m Al) in front of the DSSD all heavier products of the reaction and elastically scattered beam particles were stopped so that only light particles up to helium including elastically scattered protons reached the detector. With respect to the beam axis the DSSD covered an angle of $17{.}8^\circ~\pm~0{.}5^\circ$ at a distance of $24$~cm to the target. As usual for PSA measurements, the detector was used in reverse configuration, i.\ e.\ the n-side facing the particles. Particles punching through the DSSD were detected in a silicon PIN diode ($10\times 10$~mm$^2$, $380~\mu$m thick), which was placed behind the DSSD. With the help of this additional detector we were able to perform $\Delta$E/E measurements for comparison and, in the offline analysis, to exclude events punching through the DSSD for which the simple PSA described in this paper is not applicable.

\section{Results and discussion}
\label{results}

\begin{figure}[t]
 	\centering
	 	\includegraphics{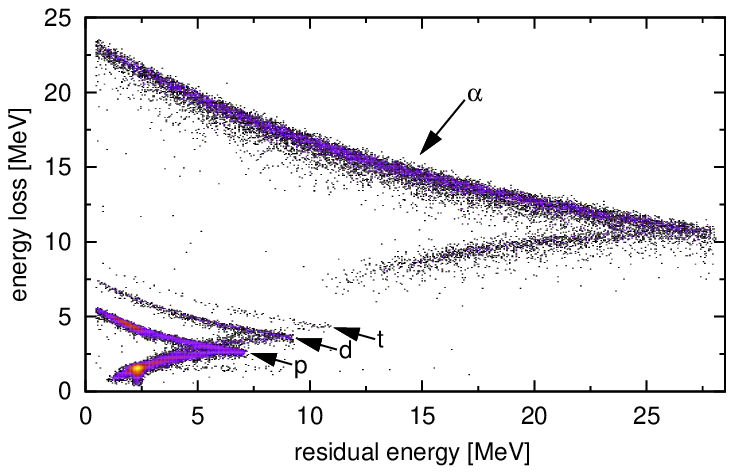}
 	\caption{Energy loss in the DSSD (energy signals taken from p-side) vs. energy of diode.}
 	\label{fig:EdeltaEp}
\end{figure}

Figure \ref{fig:TvsE} (top left) shows the measured energy plotted against the relative pulse width for strip \#3, without any further conditions on the data. The values of the pulse width plotted comprise an offset
caused by the electronics which was arbitarily set to zero for alpha particles at an energy of $23$~MeV. Despite the presence of interstrip events, alpha particles (max. energy loss of $25$~MeV in $300~\mu$m Si) can be clearly distinguished from protons~(max.\ $6{.}5$~MeV in $300~\mu$m Si).

In figure \ref{fig:TvsE} (top right), the energy measured on the p-side is plotted against the energy measured on the n-side. Events, when one single strip of each side measuring the full energy, are located on a diagonal. The off-diagonal entries arise from interstrip events. The second line appearing above the diagonal in this pattern is yet not fully understood. We assume that it is caused by particles hitting the area of the p$^+$-insulation between the strips of the n-side where the charge carriers get lost.

In figure \ref{fig:TvsE} (middle row), only those events are selected, where the particles were stopped in the DSSD, which removes the horizontal line in the PID pattern representing energy-loss signals. To exclude interstrip events, we have furthermore set a window along the diagonal in the energy-correlation plot, which is shown in figure \ref{fig:TvsE} (bottom row). The events originating from alpha particles and protons are now clearly visible in the PSA pattern and well separated from each other down to energies of about $3$~MeV. The additional weak branch comprising events with energies up to $8$~MeV, which became visible, we attribute to deuterons. Applying the condition on the interstrip events, roughly half of the events in strip~\#3 are rejected. It hast to be noted that the percentage of interstrip events is not directly connected to the geometrical interstrip area of the detector. Furthermore, the value depends on the particle energy because the extension of the field of the strips is varying with depth. This dependence can be already seen in figure \ref{fig:TvsE} (middle right) where the count rate of the off-diagonal entries is not equally distributed. For applications where the absolute detector efficiency is of importance, this effects need to be studied carefully. But of course, this is also relevant if the DSSD is used for spectroscopy only.

\begin{figure}[t]
  	\centering
  	\includegraphics{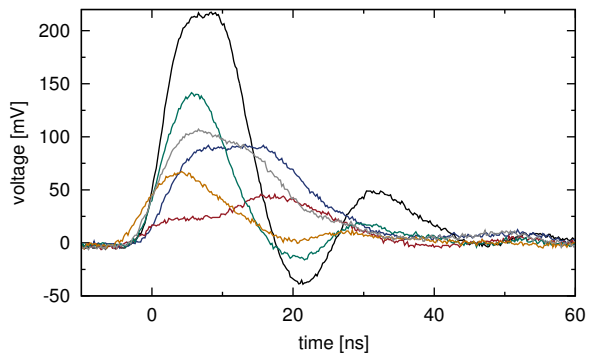}
  	\caption{Random set of pulse shapes from DSSD strip \#16 measured with a Tectronix TDS~3054C ($500$~MHz, $5$~GS/s) oscilloscope coupled, via a timing filter amplifier, to the fast timing output of the CSTA2 preamplifier.}
  	\label{fig:pulses}
	\end{figure}

For comparison, figure \ref{fig:EdeltaEp} shows a $\Delta$E/E-rest spectrum (only interstrip events are excluded) where p, d and alphas plus a few tritons can be identified. Figure \ref{fig:pulses} finally shows some pulse shapes randomly recorded with an oscilloscope connected to strip~\#16. They demonstrate the large variance of current signals in a silicon detector. The ringing, which is visible especially in the strong and fast signals, is not a property of the detector but caused by the electronics.
	
\section{Summary and conclusions}
\label{summary}

In this work the feasibility of pulse-shape analysis for particle discrimination with double-sided silicon micro-strip detectors was successfully demonstrated for the first time. Protons and alpha particles could easily be distinguished from each other down to an energy of $3$~MeV corresponding to a range of $12~\mu$m for alpha particles in silicon and $94~\mu$m for protons respectively. There is also the potential to separate deuterons and tritons for which in the applied reaction the statistics was low. Excluding the interstrip events in the offline analysis, which is an adequate procedure when using DSSDs for spectroscopy, the performance of the pulse-shape analysis is comparable to what is expected from unsegmented detectors. Motivated by these encouraging results, further research in this topic will be done. Especially the possibilities of digital sampling techniques (\cite{Bardelli:2004}, \cite{Barlini:2009}) will not only simplify the setup of electronics but could also lead to improved performance. The possibility to implement our PSA on digital electronics will be investigated for future experiments and especially for the upcoming inner shell of the EXL detector to identify p, d, t, $^3$He and alpha particles at low energies.



\section{Acknowledgments}
\label{acknowledgments}

We like to thank Reiner Kr\"{u}cken and Thomas Faestermann (Physik-Department E12, TU~M\"{u}nchen) for giving us on short notice the opportunity of having a beam time at the Tandem-van-de-Graaff accelarator at Maier-Leibnitz-Laboratorium. We also would like to thank Patrick Reichart (Universit\"{a}t der Bundeswehr M\"{u}nchen) who dispensed one day of his beam time. Another thank goes to the electronics workshop of the IKP at TU Darmstadt for their close cooperation in modifying the electronics for our purpose.

This work was supported in part by BMBF (06DA9040I) and HIC for FAIR, INTAS grant \# 06-1000012-8844, RF Federal Agency of Science and Innovations, Contract \# 03.516.11.6098 and RF President Grant \# 2951.2008.2.

\end{document}